
\magnification=1100 \vsize=24truecm \hsize=16truecm \baselineskip=0.6truecm
\parindent=1truecm \nopagenumbers \font\scap=cmcsc10 \hfuzz=0.8truecm

\null \bigskip
\centerline{\bf NON PROLIFERATION OF PREIMAGES IN INTEGRABLE MAPPINGS}
\vskip 2truecm
\centerline{\scap B. Grammaticos}
\centerline{\sl LPN, Universit\'e Paris VII}
\centerline{\sl Tour 24-14, 5${}^{\grave eme}$ \'etage}
\centerline{\sl 75251 Paris, France}
\vskip 1truecm
\centerline{\scap A. Ramani}
\centerline{\sl CPT, Ecole Polytechnique}
\centerline{\sl CNRS, UPR 14}
\centerline{\sl 91128 Palaiseau, France}
\vskip 1truecm
\centerline{\scap K.M. Tamizhmani}
\centerline{\sl Departement of Mathematics}
\centerline{\sl Pondicherry University}
\centerline{\sl 605014 Pondicherry, India}
\bigskip
\bigskip
\noindent Abstract \smallskip \noindent We present an integrability
criterion for rational mappings based on two requirements. First, that
a given point should have a unique preimage under the mapping and,
second, that the spontaneously appearing singularities be confined to
a few iteration steps. We present several examples of known integrable
mappings that meet these requirements and, also, use our algorithm in
order to derive new examples of integrable mappings. \vfill\eject

\footline={\hfill\folio} \pageno=2
{\scap 1. Introduction} \medskip
\noindent Integrability detectors are rare even for continuous systems.
Recent progress in the integrability of discrete systems has spurred
the activity in this direction, leading to interesting results of
great variety. In order to characterize discrete integrability, Arnold
has introduced and investigated the concept of complexity for mappings
in a plane [1]. Arnold defines the complexity from the number of
intersection points of a fixed curve with the image of a second given
curve under the $k^{\rm th}$ iteration of the mapping. For a
polynomial mapping the growth of the number of intersection points is
in general exponential in $k$. However for integrable mappings the
growth is only polynomial in $k$. This result is included in a more
general analysis presented by Veselov [2] discussing the dynamics of
multiple-valued mappings (correspondences) and the growth of the
number of different images (and preimages). Veselov, too, has linked
integrability to ``slow growth''.

Our approach of discrete integrability is different from the above
since it was based, essentially, on rational (rather than polynomial)
mappings. For rational mappings, an important question is what happens
whenever accidentally (i.e. depending on the initial conditions) a
denominator vanishes, leading to a divergent mapping variable. In
general one expects this singularity to propagate indefinitely under
the mapping iterations, but it turns out that for {\sl integrable}
mappings these singularities disappear after a few steps. This
observation has led to the proposal of the singularity confinement [3]
criterion as detector of discrete integrability. Its efficiency has
already been proven through the derivation of new integrable systems
leading to the discovery of discrete Painlev\'e equations [4].

It appears that the two notions of {\sl slow growth} and {\sl confined
singularities} play an important role in the characterization of the
integrable discrete systems.  In what follows, we will try to present
our approach which is based on both notions.  We will, first,
introduce the notion of preimage nonproliferation as well as the
algorithm for its assessment. Based on the slow-growth principle, we
claim that the number of preimages of a given point should not grow
exponentially fast, which, when we consider mappings rather than
general correspondences, can only mean a single preimage. Therefore,
for the practical implementation we will present of this criterion, we
will require that the inverse of the mapping be uniquely defined. The
singularity confinement conjecture will also be extended in the
following sections, essentially through the extension of the notion of
singularity of a mapping. Apart from the singularity related to a
divergence, we will consider as appearance of a singularity all the
instances where the mapping accidentally loses some degrees of
freedom. (The precise mechanism will become clear in Section
3). Confinement of this singularity consists in the recovery of these
lost degrees of freedom usually through the appearance of an
indeterminate form like $0/0$. While preimage nonproliferation is only
a necessary condition for integrability, we conjecture that its
combination with singularity confinement leads to a sufficient
condition for integrability of discrete systems.

In what follows, we will limit ourselves to rational explicit
mappings, i.e.  $$x_i'=f_i(x_1,x_2,\dots x_N)\quad\quad\quad\quad
i=1,\dots N \eqno(1)$$ with rational $f_i$'s. Integrability in our
sense means one of the following things:

\noindent a) existence of a sufficient number of rational
$\Phi_k(x_1,\dots x_N)=C_k$ the values of which are invariant under
the action of the mapping.

\noindent b) linearizability of the mapping through a Cole-Hopf
type transformation $x_i={P_i/Q_i}$ whereupon the mapping reduces to a
linear one for the $P_i$'s, $Q_i$'s.

\noindent c) linearizability through a Lax pair.
In this case, the mapping is the compatibility condition of a linear
system of differential-difference, q-difference or pure difference
equations.

The above are not definitions but rather illustrations of the various
types of integrability. It may well occur, as in the case of Quispel's
mappings [5], that the existence of one invariant reduces the mapping
to a correspondence of the form $F(x,x')=0$ that can be parametrized
in terms of elliptic functions. In other cases, integration using the
rational invariants may lead to some transcendental equation like the
discrete Painlev\'e ones. All of the above types of integrability have
been encountered in the discrete systems that we have studied
[6,7]. The reason for the above classification is to emphasize the
parallel existing between the continuous and discrete cases.  In the
next section , we examine specific examples of mappings and formulate
along the way our conjecture on preimage nonproliferation.  \bigskip
{\scap 2. Examples of integrable mappings and the preimage
nonproliferation criterion}
\medskip
\noindent Let us start with a very simple
example of rational mapping, in which the growth of the number of
preimages must be invoked. In [6], we studied the one-component,
two-points mapping of the form: $$x'=f(x) \eqno(2)$$ where $f$ is
rational. Singularity confinement considerations lead to $$f(x)=
\alpha + \sum_k {1 \over (x-\beta_k)^{\nu_k}} \eqno(3)$$ with integer
$\nu_k$, provided that for all $k$, $\beta_k\neq \alpha$. Indeed, if
$x=\beta_k$ at some step, then $x'$ diverges, $x''=\alpha$ and $x'''$
is finite.  So the mapping propagates without any further
difficulty. However, if we consider the ``backward'' evolution, then
(2) solved for $x$ in terms of $x'$ leads to multideterminacy and the
number of preimages grows exponentially with the number of
``backward'' iterations. Indeed, the only mapping of the form (2-3)
with no growth is just the homographic: $$x'={a x + b \over cx + d}
\eqno(4)$$ which is the discrete form of the Riccati equation.Thus, in
this case, the argument of slow growth of the number of preimages of
$x$ is essential in deriving the form of the discrete Riccati
equation.

Another classical example in the domain of integrable mappings is the
Quispel familly. In [5], Quispel and collaborators have shown that the
mappings $${\overline x}= {f_1(y)-f_2(y)x\over f_2(y)-f_3(y)x}$$
$$\eqno(5)$$ $${\overline y}={g_1({\overline x})-g_2({\overline
x})y\over g_2({\overline x})-g_3({\overline x})y}$$ are integrable,
provided the $f_i$'s, $g_i$'s are specific quartic polynomials
involving 18 parameters. We remark here that the mapping is
``staggered'', i.e. while ${\overline x}$ is defined in terms of
$(x,y)$, ${\overline y}$ is defined in terms of $({\overline
x},y)$. It is precisely this staggered structure that allows one to
define a unique preimage to $(x,y)$. As Quispel has shown in [8], the
mapping (5) is reversible which means that it can be written as a
product of two involutions. It is not clear whether reversibility is a
prerequisiste for integrability but though reversibility ensures that
the preimage is unique, still there exist reversible systems that are
not integrable.

Reversible integrable mappings have also been considered by the Paris
group [9] in their works based on the study of lattice spin and vertex
models. They have shown that the transformations involved are in fact
symmetries of the Yang-Baxter equations. These symmetries are
constructed as the product of a pair of noncommuting involutions: thus
the mapping is reversible and generically of infinite order. Still, it
is interesting to investigate the mechanism for the nonproliferation
of preimages in this case and present the algorithm that one should
use. Let us illustratre this in the case of the mapping:
$$x'={x+y-2xy^2\over y(y-x)}$$ $$\eqno(6)$$ $$y'={x+y-2yx^2\over
x(x-y)}$$ The first step consists in considering the system of $N$
equations $x'_i-f_i(x_k)=0$ and eliminate successively all the
$x'_i$'s but one. In the case of the mapping (6) the resultant in $x$
after eliminating $y$ (and vice-versa) is a fifth-degree polynomial in
$x$ (resp. $y$) with coefficients depending on $x'$ and $y'$. Next we
factorize this resultant. Preimage nonproliferation requires that only
{\it one} factor depend on $x', y'$, the other factors being
associated to indeterminate forms $0/0$. We find in the particular
example the factors $x^2(x^2-1)$ (resp. $y^2(y^2-1)$) and one last
factor leading to $$x={x'-y'\over y'^2+x'y'-2} $$ $$\eqno(7)$$
$$y={y'-x'\over x'^2+x'y'-2}$$ This is the typical situation for
integrable rational mappings. The factorization of the resultant gives
the unique inverse of the mapping along with particular values (here
$x=y=0,\pm 1$) corresponding to the indeterminate forms of the
mappings.

Veselov has studied the integrability of polynomial mappings and has
shown [2] that the mapping $x'=P(x,y),y'=Q(x,y)$ is integrable (in the
sense that it possesses a nonconstant polynomial integral $\Phi
(x,y)$) if there exists a polynomial change of coordinate variables
transforming the mapping to triangular form: $$x'=\alpha x +P(y)$$
$$y'=\beta y+\gamma\eqno(8)$$ for polynomial $P$. Moreover he has
shown that in this case the complexity of the mapping is bounded. The
important feature in (8) is the fact that the equation for $y'$ is
linear. Thus the inversion of (8) is straightforward. Thanks to the
triangular form the integration of (8) is reduced to the solution of
two affine mappings, first for $y$ and then for $x$.

One more interesting illustration of the preimage nonproliferation
algorithm is provided by the discrete Painlev\'e equations that we
derived in [10] and which are {\it not} of Quispel form.  We have
found there that the mappings: $$x'={xy(a(y+1)-xy^2)\over a(y+1)^2}$$
$$\eqno(9)$$ $$y'={a(y+1)(xy^2-(y+1)(a-zy)\over (a(y+1)-xy^2)^2}$$
where $a=cnst.$ and $z$ linear in the lattice variable, is a discrete
form of the P$_{\rm{I}}$ equation. In order to check the preimage
nonproliferation, we eliminate $x$ (or $y$) from (9) and factorize the
resultant. We find as expected factors related to pathological points
$x=0$, $y=0$, $y=-1$, and a unique inverse that reads:
$$x={x'(x'y'+z)[(x'y'+z)^2+a(y'+1)(z-x')]\over a(y'+1)^2}$$
$$\eqno(10)$$ $$y={a(z-x')(y'+1)\over (x'y'+z)^2}$$ Thus in this
example, too, as in all previous ones, integrability is related to
non-growth of the number of preimages.
\bigskip

{\scap 2. Extending the singularity confinement criterion}
\medskip
\noindent In the previous section, we encountered several examples
of integrable mappings, all of which satisfied the ``no-growth''
property. Here, we will apply the preimage nonproliferation criterion
in order to construct explicitely integrable mappings. However since
this criterion furnishes only a necessary condition for integrability,
we will supplement it by singularity confinement, our conjecture being
that the combination of the two criteria is sufficient for discrete
integrability.

Applying the preimage nonproliferation algorithm to a general mapping
can easily lead to untractable calculations. If, however, there are
not too many free parameters in the mapping the implementation of the
criterion is straightforward. In what follows, we will limit ourselves
to simple two-component, two-point mappings of the form
$$x'={Q_1(x,y)\over Q(x,y)}\quad\quad y'={Q_2(x,y)\over
Q(x,y)}\eqno(11)$$ where $Q$, $Q_1$ and $Q_2$ are quadratic
polynomials in $x$ and $y$. By applying a general linear
transformation on this mapping we can reduce the (common) denominator
to one of the two canonical forms $Q=xy-1$ or $x^2-y$ (or any of the
degenerate forms $Q=xy$, $x^2-1$ or $x^2$). Let us start with the
mapping $h$:
$$x'={a_{20}x^2+a_{11}xy+a_{02}y^2+a_{10}x+a_{01}y+a_{00}\over
xy-1}\eqno(12.a)$$
$$y'={b_{20}x^2+b_{11}xy+b_{02}y^2+b_{10}x+b_{01}y+b_{00}\over
xy-1}\eqno(12.b)$$ One assumption that we will introduce here is
$a_{02}=0$ since it leads to a great simplification of the
calculations. As a first step in the preimage nonproliferation
algorithm we eliminate $y$ between (12.a-b) for given $x'$ and $y'$
and obtain a resultant that is a fourth degree polynomial in $x$. We
demand that three of the roots be independent of $x'$, $y'$ and denote
them by $x_1$, $x_2$ and $x_3$. We then demand that whenever $x=x_i$,
$y=1/x_i$, $i=1,2,3$ both $x'$ and $y'$ have the indeterminate form
$0/0$. Calling: $$\Sigma = x_1+x_2+x_3$$ $${\rm P} = x_1 x_2+x_2
x_3+x_3 x_1 \eqno(13)$$ $$\Pi = x_1 x_2 x_3$$ we find as a condition
for a unique preimage: $$a_{10}=-a_{20}\Sigma$$ $$a_{01}=-a_{20}\Pi$$
$$a_{00}=a_{20}{\rm P}-a_{11} \eqno (14)$$
$$b_{10}=-b_{20}\Sigma-b_{02}/\Pi$$ $$b_{01}=-b_{20}\Pi-b_{02}{\rm
P}/\Pi$$ $$b_{00}=b_{20}{\rm P}-b_{11}+b_{02}\Sigma/\Pi $$ With (14)
the mapping satisfies the preimage nonproliferation requirement. This
is {\it not}, however, sufficient for integrability. What we must also
demand is that the mapping have confined singularities. The simplest
kind of singularity is whenever the denominator vanishes,
i.e. $Q(x,y)=0$ or in our example $y=1/x$. However, since in the
present case the numerators $Q_1$, $Q_2$ are also quadratic, the
singularity is confined in one step: $Q(x,y)=0$ leads to diverging
$x'$, $y'$ and because the degrees of numerators and (common)
denominator are equal this leads to finite $x''$ and $y''$. So the
study of this singularity does not introduce any constraint on the
mapping. But the vanishing of the denominators is {\it not} the only
singularity of the mapping: a subtler singularity may exist.

Normally for a general $N$-component mapping, $N$ free parameters,
introduced by the initial conditions, must be present at every
step. Now, it may happen that at some iteration one (or more) degress
of freedom be lost.  The condition for this to occur is that the
Jacobian of $(x'_1,x'_2,\dots x'_N)$ with respect to $(x_1,x_2,\dots
x_N)$ vanishes. For a general mapping $x'_i=f_i(x_k)$ this reads:
$$J=\left|\matrix{ {\partial x'_1\over\partial x_1}&{\partial
x'_1\over\partial x_2}&\ldots& {\partial x'_1\over\partial x_N}\cr
{\partial x'_2\over\partial x_1}&{\partial x'_2\over\partial
x_2}&\ldots& {\partial x'_2\over\partial x_N}\cr
\vdots &\vdots &\ddots &\vdots \cr
{\partial x'_N\over\partial x_1}&{\partial x'_N\over\partial
x_2}&\ldots& {\partial x'_N\over\partial x_N}\cr}\right| =0\eqno(15)$$
How can this singularity be confined? By this we mean that the mapping
must recover the lost degree of freedom. For rational mapping of the
kind we are considering, this can be realized if some of the mapping's
variables assume an indeterminate form $0/0$. In that case new free
parameters can be introduced and the mapping recovers its full
dimensionality.

Let us apply this criterion to the mapping $h$. The Jacobian readily
factorizes and we obtain three factors: $$x+x_i-\Sigma+{y\Pi/x_i}=0
\quad\quad i=1,2,3 \eqno(16)$$ Thus whenever (16) is satisfied a
singularity appears (in the sense of the loss of one degree of
freedom). For the confinement of this singularity (at the $x''$, $y''$
level) we must have $x'y'-1=0$ and the numerators of both $x''$ and
$y''$ must vanish. First we supplement the condition for the vanishing
denominator $x'y'-1$. This leads to a number of equations that, in
fact, specify fully the remaining $a$, $b$ coefficients and moreover
put a constraint on $x_1,x_2,x_3$. The latter can be written (up to an
odd permutation of $x_1,x_2,x_3$) as:
$$3x_1x_2x_3=x_1^2x_3+x_2^2x_1+x_3^2x_2 \eqno(17)$$ For the $a,b$ we
obtain: $$a_{11}^3=-\Pi$$ $$a_{20}=-{1\over a_{11}}$$
$$b_{02}={a_{11}}\eqno(18)$$ $$b_{20}={1\over \Pi}$$ $$b_{11}={{\rm
P}\over \Pi}-{1\over a_{11}}$$ In fact the simplest way to parametrize
equations (14,17,18) is to take $a_{11}\equiv a$ as basic
parameter. Introducing one further parameter $\omega$ we can express
$x_1,x_2,x_3$ as: $x_1=-a\omega$, $x_2=a(1+1/\omega)$ and
$x_3=a/(1+\omega)$.  It turns out that once conditions (14,17,18) are
implemented the numerators of both $x''$ and $y''$ automatically
vanish. Thus the mapping $h$ is singularity confining and according to
our conjecture it should be integrable. This is indeed the case and
one invariant can easily be found. It reads:
$$\prod_{i=1}^3{x+x_i-\Sigma+y\Pi /x_i\over x-x_i}=(-1)^nK\eqno(19)$$
i.e. the product on the lhs, instead of being strictly constant,
alternates sign between even and odd iterations. One should, in
principle, take the square of the lhs in order to find a true
constant. A closer inspection of the mapping (motivated by the form of
the invariant) reveals an even simpler structure: the mapping is
periodic with period six, i.e.  $h^6=I$.

Finaly, the mapping can be cast in a much simpler form if one use the
scaling freedom in order to reduce the number of the parameters from
two to one. We shall not enter into these details but we just give the
final result: $$x'={-x^2+xy+\sigma x-y+2-\sigma\over
xy-1}\eqno(20.a)$$
$$y'={-x^2+(2-\sigma)xy+y^2+(1+\sigma)x+(\sigma-4)y+1-\sigma\over
xy-1}\eqno(20.b)$$

In an analogous way we can treat the mapping $p$:
$$x'={a_{20}x^2+a_{11}xy+a_{02}y^2+a_{10}x+a_{01}y+a_{00}\over
x^2-y}\eqno(21.a)$$
$$y'={b_{20}x^2+b_{11}xy+b_{02}y^2+b_{10}x+b_{01}y+b_{00}\over
x^2-y}\eqno(21.b)$$ with $a_{02}=0$.  As in the previous case, we can
ask that the resultant of the elimination of $y$ between (21.a-b) for
given $x'$, $y'$ (which is quartic in $x$), possess three roots
independent on $x'$, $y'$, namely $x_1$, $x_2$ and $x_3$,
corresponding to $0/0$ indeterminacies.  In a second step, the
singularity confinement can be implemented in a perfect parallel to
the case of the mapping $h$, leading to: $$x'={x^2\Sigma-xy- x{\rm
P}+\Pi\over x^2-y}\eqno(22.a)$$
$$y'={x^2\Sigma^2-2xy\Sigma+y^2-(\Sigma{\rm P}+\Pi)x+y{\rm
P}+\Sigma\Pi\over x^2-y}\eqno(22.b)$$ with $\Sigma ,{\rm P},\Pi$ given
by (13).  As in the previous case, the freedom of transformations can
be used in order to simplify this mapping. Finally only one parameter
remains and the mapping reads: $$x'={x(x-y-\rho)\over
x^2-y}\eqno(23.a)$$ $$y'={(x-y)(x-y-\rho)\over x^2-y}\eqno(23.b)$$
This mapping is indeed integrable, but in a trivial way: it is just an
involution, $p^2=I$.  Still, the important point here is that the
conjecture concerning the integrability of mappings that have
nonproliferating preimages and confined singularities is once more
satisfied.
\bigskip
{\scap Conclusion}
\medskip
\noindent In the preceding sections, we have presented in detail the
preimage nonproliferation criterion, which, we conjecture, is a
necessary condition for the integrability of rational mappings. Based
on the ``slow-growth'' principle, this criterion consists in requiring
that, in order to be a candidate for integrability, a rational mapping
possess a unique inverse. Thus the number of preimages of a given
point through the mapping does not grow with the number of iterations
(while an exponential increase is, generically, expected).  Since this
criterion offers only necessary conditions it cannot predict
integrability but can be used as a fast ``screening'' procedure. The
successful candidates can then be tested for singularity confinement,
which is more stringent but of more difficult implementation. The
combination of the two criteria (preimage nonproliferation and
singularity confinement) we conjecture to be an integrability
predictor of the same efficiency as the Painlev\'e method for
continuous systems. Several results exist already based on this
approach and we expect the extension of the singularity confinement
presented here to further widen its range of applications.
\bigskip
{\scap Acknowledgements}.
\medskip
\noindent K.M Tamizhmani is grateful to Paris VI and VII Universities,
the International Mathematical Union and the French Ministry of
Foreign Affairs for the invitation and financial assistance that made
possible the present collaboration.
\bigskip
{\scap References}.
\medskip
\item{[1]} V.I. Arnold, Bol. Soc. Bras. Mat. 21 (1990) 1.
\item{[2]} A.P. Veselov, Comm. Math. Phys. 145 (1992) 181.
\item{[3]} B. Grammaticos, A. Ramani and V. Papageorgiou, Phys. Rev.
Lett. 67 (1991) 1825.
\item{[4]} A. Ramani, B. Grammaticos and J. Hietarinta, Phys. Rev.
Lett. 67 (1991) 1829.
\item{[5]} G.R.W. Quispel, J.A.G. Roberts and C.J. Thompson, Physica
D34 (1989) 183.
\item{[6]} A. Ramani, B. Grammaticos and G. Karra, Physica A 81 (1992)
115.
\item{[7]} V.G. Papageorgiou, F.W. Nijhoff, B. Grammaticos and A.
Ramani, Phys.  Lett. A164 (1992) 57.
\item{[8]} G.R.W. Quispel and J.A.G. Roberts, Phys. Lett. A132 (1988)
161.
\item{[9]} M. Bellon, J-M. Maillard and C-M. Viallet, Phys. Lett.
A159 (1991) 221 and 233.
\item{[10]} A.S. Fokas,  A. Ramani, B. Grammaticos, J. Math. An.
Appl.  to appear.

\end